\documentclass[preprint,showpacs,superscriptaddress]{revtex4-1}

\usepackage{graphicx}
\usepackage{dcolumn}
\usepackage{bm}

\usepackage{color}

\begin{document}


\title{Surface Kondo Effect and Non-Trivial Metallic State of the Kondo Insulator YbB$_{12}$}

\author{Kenta Hagiwara}
\affiliation{Department of Physics, Graduate School of Science, Osaka University, Toyonaka 560-0043, Japan}
\author{Yoshiyuki Ohtsubo}
\email{y\_oh@fbs.osaka-u.ac.jp}
\affiliation{Department of Physics, Graduate School of Science, Osaka University, Toyonaka 560-0043, Japan}
\affiliation{Graduate School of Frontier Biosciences, Osaka University, Suita 565-0871, Japan}

\author{Masaharu Matsunami}
\author{Shin-ichiro Ideta}
\author{Kiyohisa Tanaka}
\author{Hidetoshi Miyazaki}
\affiliation{UVSOR Facility, Institute for Molecular Science, Okazaki 444-8585, Japan}

\author{Julien Rault}
\author{Patrick Le F\`evre}
\author{Fran\c{c}ois Bertran}
\affiliation{Synchrotron SOLEIL, Saint-Aubin-BP 48, F-91192 Gif sur Yvette, France}
\author{Amina Taleb-Ibrahimi}
\affiliation{Synchrotron SOLEIL, Saint-Aubin-BP 48, F-91192 Gif sur Yvette, France}
\affiliation{UR1/CNRS-Synchrotron SOLEIL, Saint-Aubin, F-91192 Gif sur Yvette,
France}

\author{Ryu Yukawa}
\author{Masaki Kobayashi}
\author{Koji Horiba}
\author{Hiroshi Kumigashira}
\affiliation{Photon Factory, Institute of Materials Structure Science, High Energy Accelerator Research Organization (KEK), 1-1 Oho, Tsukuba 305-0801, Japan}

\author{Fumitoshi Iga}
\affiliation{College of Science, Ibaraki University, Mito 310-8512, Japan}

\author{Shin-ichi Kimura}
\email{kimura@fbs.osaka-u.ac.jp}
\affiliation{Department of Physics, Graduate School of Science, Osaka University, Toyonaka 560-0043, Japan}
\affiliation{Graduate School of Frontier Biosciences, Osaka University, Suita 565-0871, Japan}

\date{\today}

\begin{abstract}
A synergistic effect between strong electron correlation and spin-orbit interaction (SOI) has been theoretically predicted to result in a new topological state of quantum matter on Kondo insulators (KIs), so-called topological Kondo insulators (TKIs).
One TKI candidate has been experimentally observed on the KI SmB$_6$(001), and the origin of the surface states (SS) and the topological order of SmB$_6$ has been actively discussed.
Here, we show a metallic SS on the clean surface of another TKI candidate YbB$_{12}$(001), using angle-resolved photoelectron spectroscopy.
The SS showed temperature-dependent reconstruction corresponding with the Kondo effect observed for bulk states.
Despite the low-temperature insulating bulk, the reconstructed SS with $c$-$f$ hybridization was metallic, forming a closed Fermi contour surrounding $\bar{\Gamma}$ on the surface Brillouin zone and agreeing with the theoretically expected behavior for SS on TKIs.
These results demonstrate the temperature-dependent holistic reconstruction of two-dimensional states localized on KIs surface driven by the Kondo effect.
\end{abstract}

\maketitle


Recently, non-trivial surface electronic structures such as surface metallic states of topological insulators \cite{Kane10, Qi11} and giant Rashba-type spin splitting of polar semiconductors \cite{Ishizaka11, Crepaldi12, Landolt12} have been observed.
These electronic structures originate from the large spin-orbit interaction (SOI) of heavy elements.
On the other hand, in metallic compounds containing heavy elements, especially rare-earths, heavy quasiparticles, namely heavy fermions, appear owing to the Kondo effect \cite{Takabatake98}.
The origin of the Kondo effect is a hybridization between the conduction band and localized states ({\it e.g.} 4f states of rare-earths), namely, $c$-$f$ hybridization, derived from the Anderson model.
At the surface of rare-earth intermetallic compounds, new physical properties originating from the Kondo effect and the large SOI such as superconductivity without inversion symmetry \cite{Bauer04} will appear.
Kondo insulators (KIs) are one candidate of such compounds.

In general, KIs possess a small energy gap (typically a few tens of meV) at the Fermi level ($E_F$) owing to $c$-$f$ hybridization at low temperatures \cite{Kumigashira01}.
In the case that the $c$-$f$ gap is formed by the conduction and valence bands with inverted parities, two-dimensional metallic electronic states should always appear on the KI surface as in the case of topological insulators \cite{Dzero10, Takimoto11}.
Such materials are categorized as topological Kondo insulators (TKIs) in which the topological surface states originate from {\it c-f} gap formation owing to the Kondo effect.
Thus, TKIs are new physical states of quantum matter driven by the synergistic effect between strong electron correlation and SOI.

The surface state of the KI samarium hexaboride (SmB$_6$) has been investigated both theoretically and experimentally \cite{Miyazaki12, Jiang13, Denlinger14, Hlawenka15} and has been theoretically predicted to be a non-trivial topological surface state of a TKI. This predicted surface electronic structure has been experimentally observed \cite{Jiang13, Denlinger14}.
However, the origin of the metallic surface state is currently under debate because Hlawenka {\it et al.} recently reported that the surface metallic state trivially originates from large Rashba splitting \cite{Hlawenka15}.
Therefore, a survey of another material is desirable to provide further insight into the origin of metallic surface states on KIs.

Ytterbium dodecaboride (YbB$_{12}$) is a typical KI which has a NaCl-type crystal structure with Yb and B$_{12}$ clusters as shown in Fig. 1(a) \cite{Iga98}.
A clear energy gap appears in the bulk of YbB$_{12}$ with a gap size of about 40~meV of the peak (15~meV of the onset) which has been observed by an optical conductivity measurement \cite{Okamura05}.
Using ``angle-integrated'' photoemission measurements, a ``pseudo''-gap opens at $E_F$, but a finite density of states (DOS) has been observed at temperatures lower than that of the full gap opening \cite{Okawa15}.
The observed finite DOS at $E_F$ is considered to originate from a metallic surface state, and this metallic surface state has been confirmed using electrical transport measurements \cite{Iga_Priv}.
Moreover, a theoretical study has predicted that this metallic surface conductivity originates from topological surface states \cite{Weng14}.
However, the band structure of YbB$_{12}$ has not been observed using momentum-resolved measurements such as angle-resolved photoelectron spectroscopy (ARPES) because a well-defined clean surface has not been obtained \cite{Takeda06}.
We recently succeeded in obtaining a clean YbB$_{12}$ surface.
Here, we show the first well-defined ARPES data of YbB$_{12}$ and discuss the origin of the metallic surface state.

In this article, we report a surface state (SS) on a clean surface of the KI YbB$_{12}$(001) and its temperature-dependent reconstruction based on ARPES.
The state was metallic and showed no dispersion along $k_z$, indicating its surface localization.
While the SS does not hybridize with the Yb $4f$ state lying immediately below the Fermi level ($E_{\rm F}$) at room temperature, strong hybridization occurred at low temperatures.
The reconstructed SS due to the $c$-$f$ hybridization at 20 K was metallic and was continuously dispersed across the bulk bandgap of the KI between $E_{\rm F}$ and the $\sim$50-meV binding energy.
This low-temperature SS behaviour agrees with the expected behaviour for SS on TKIs.
Moreover, these results demonstrate the temperature-dependent holistic reconstruction of two-dimensional states localized on the KI surface driven by the Kondo effect.

\subsection*{Characterization of the YbB$_{12}$(001) clean surface}
Figure 1 (b) shows the low-energy electron diffraction (LEED) pattern observed after the cleaning process (see Methods).
As shown by the sharp spots and low backgrounds, a well-ordered clean YbB$_{12}$(001) surface was obtained.
In addition to the integer order spots corresponding to a (001) in-plane lattice constant (5.28 \AA), we found fractional order spots showing $c$(2$\times$2) surface periodicity (Fig. 1 (b)).
The LEED pattern shows fourfold rotation symmetry, which is expected from the bulk crystal structure (Fig. 1 (a)).

Figure 1 (c) shows the wide-valence band spectrum for the YbB$_{12}$(001)-$c$(2$\times$2) surface taken at h$\nu$ = 80 eV.
The Yb$^{2+}$ and Yb$^{3+}$-$4f$ levels are observed at similar energy positions to those measured with scraped or cleaved YbB$_{12}$ single crystals \cite{Susaki96, Takeda06}.
The major difference observed in this work is the absence of Yb$^{2+}$ components around 0.9 and 2.2 eV, which are assigned to be ``surface''  components.
It would be because the Yb atoms at the topmost surface are desorbed during the heating process.
Indeed, as shown in Fig. 1(d), we found a clear surface component at the B-$1s$ level at the binding energy of 188 eV with a surface-sensitive condition ($h\nu$ = 280 eV, kinetic energy $\sim$90 eV), but this appeared as a weak tail in the bulk-sensitive measurement ($h\nu$ = 1000 eV).
Regarding the Yb$^{2+}$-$4f$ levels, Fig. 1(e) shows almost the same peak positions with surface/bulk sensitive conditions, supporting the above assumption that Yb atoms are not located at the topmost surface layer but in the deeper layers. 

\begin{figure}[p]
\includegraphics[width=80mm]{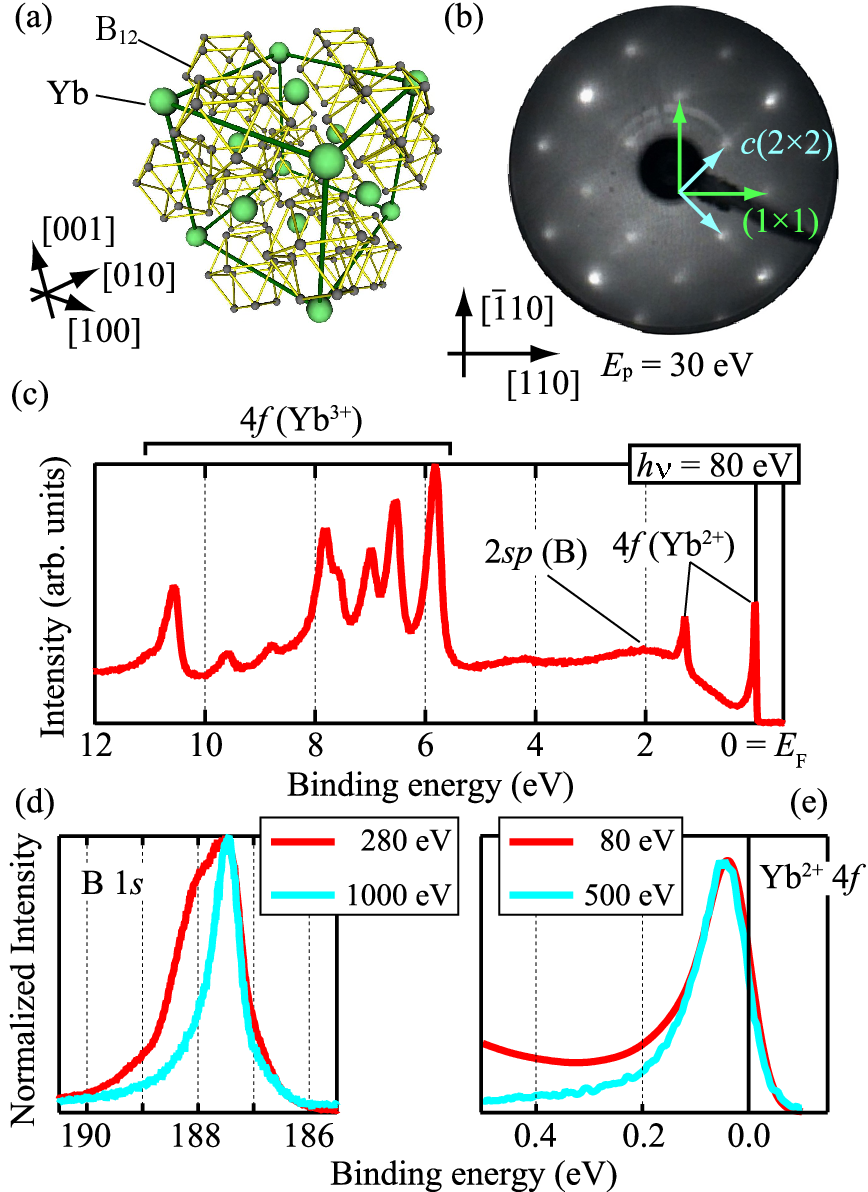}
\caption{\label{fig1}
(a) Crystal structure of YbB$_{12}$ \cite{Saso_Crystal}.
(b) A low-energy electron diffraction pattern of the clean YbB$_{12}$(001) surface at room temperature. Arrows indicate the surface unit vectors.
(c--e) Angle-integrated photoelectron spectra taken at 20 K with photon energies at (c) 80, (d) 280/1000, and (e) 80/500 eV. In (e), the spectrum taken at 80 is convolved with a Gaussian (full-width at half-maximum of 75 meV) for comparison with other spectra at 500 eV taken with lower energy resolution.
}
\end{figure}



\begin{figure}[p]
\includegraphics[width=80mm]{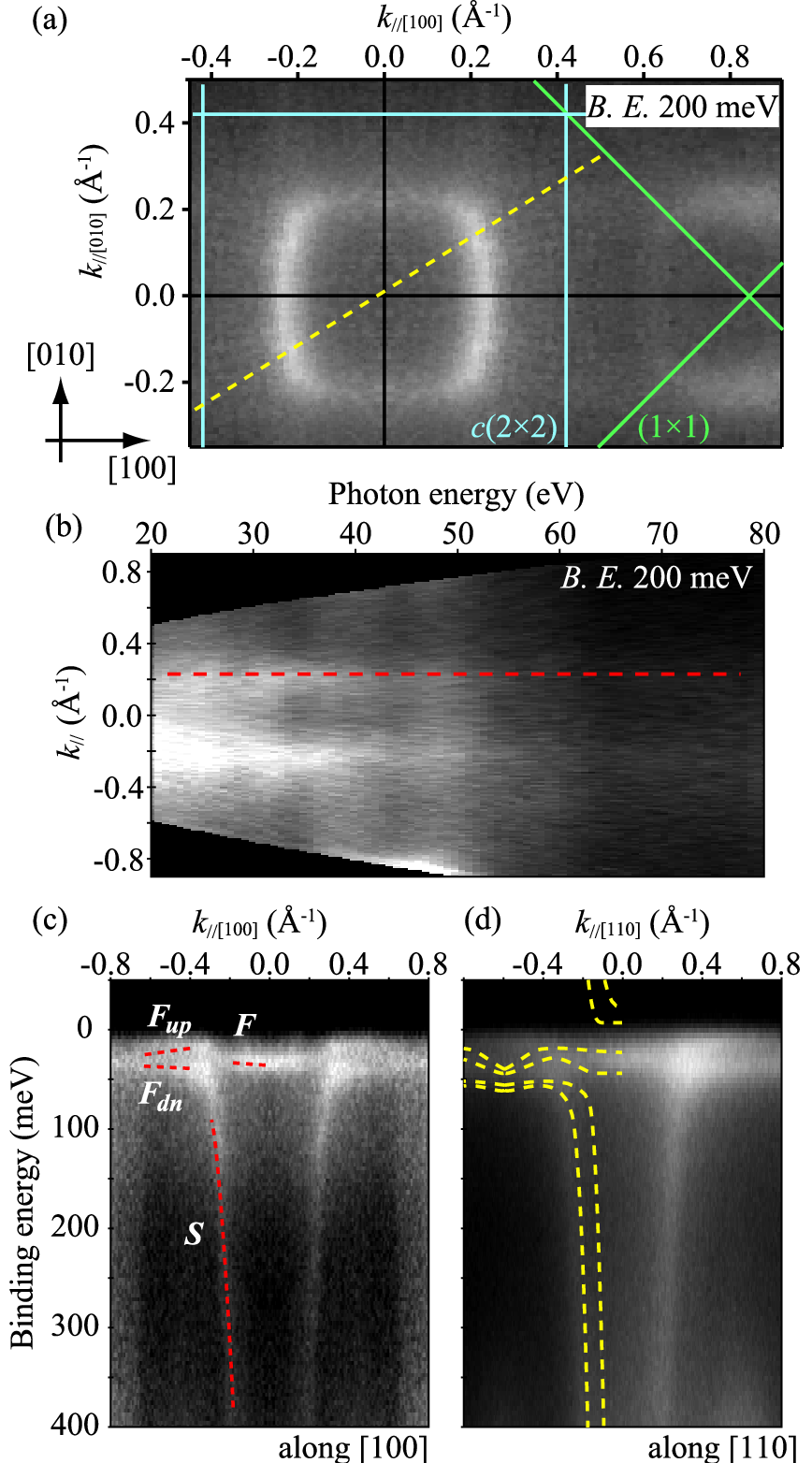}
\caption{\label{fig2}
Angle-resolved photoelectron spectroscopy (ARPES) data taken below 20 K.
(a) Constant energy contour at the binding energy of 200$\pm$10 meV ($h\nu$ = 50 eV). Thin lines represent the surface Brillouin zone (BZ) boundary. Arrows indicates the Miller indices based on the bulk crystal structure. The indexes used in the following parts obey these definitions.
(b) Momentum distribution curves along a dashed line in (a) at the binding energy of 200$\pm$10 meV as a function of photon energies. The dashed line provides a guide to the eye.
(c, d) ARPES intensity maps along (c) [100] and (d) [110] measured with $h\nu$ = 53.5 eV. 
Dashed lines in the left region of (c) indicate the observed bands named $S$, $F$, and $F_{dn/up}$.
Those in (d) are calculated bulk bands around the $X$ point of bulk BZ captured from ref. \cite{Weng14}.
}
\end{figure}

\subsection*{Surface states on YbB$_{12}$(001)}

Figure 2 shows ARPES data taken below 20 K.
As shown in Fig. 2 (a), square constant energy contours (ECs) are observed at the binding energy of 200 meV.
These contours fold with respect to the $c$(2$\times$2) surface Brillouin zone (SBZ) boundary, suggesting they originate from the surface.
Figure 2 (b) shows the intensity plots of the momentum distribution curve (MDCs) along the dashed line in Fig. 2 (a) as a function of probing photon energies.
As indicated by the dashed line, the MDC peak corresponding to the square EC does not change its in-plane wavevector, indicating that this state is a two-dimensional surface state.

Figures 2 (c) and (d) are the band dispersions along [100] and [110], respectively. In what follows, we use the Miller indices based on the bulk crystal structure to define the direction in reciprocal space.
The surface state ($S$ in Fig. 2 (c)) identified above shows a steep dispersion below 100 meV.
Dashed lines superposed on Fig. 2 (d) are the calculated bulk bands based on the local density approximation (LDA) plus Gutzwiller method \cite{Weng14}.
We captured the bulk bands around the $X$ point of bulk Brillouin zone (BZ).
From the calculated bulk bands along $\Gamma$-$X$ (see ref. \cite{Weng14}), the highly dispersive conduction band clearly has its minimum at $X$; this means the bulk band dispersion around $X$ is almost at the lower edge of the projected bulk bands in SBZ around $\bar{\Gamma}$ onto which the $\Gamma$-$X$ line is projected.
As shown, the $S$ band dispersion is almost parallel to the bulk band.
Note that $S$ shows no dispersion along $k_z$, which clearly differs from the bulk conduction band dispersing along the surface normal (see ref. \cite{Weng14}).
Such SS dispersion along bulk bands are known for so-called Shockley-type SS on noble metal surfaces \cite{Kevan87} and for surface resonances localized in the subsurface region of semiconductor surfaces \cite{Ohtsubo13}.
This type of SS originates from bulk bands, but it is localized in the surface/subsurface region because of the truncation of the three-dimensional (3D) periodicity of the crystal.

In addition to the highly dispersive $S$ band, there is a less dispersive state around 30 meV ($F$ in Fig. 2 (c)), probably originating from the Yb$^{2+}$-4$f$ bands.
The $F$ band appears as single peak at $\bar{\Gamma}$ ($F_0$) but splits to two branches ($F_{up}$ and $F_{dn}$) at $|k| >$ 0.4 \AA$^{-1}$.
These states cross with the $S$ band around 0.3 \AA$^{-1}$ where they apparently hybridize with each other.
Such behaviour suggests $c$-$f$ hybridization driven by the Kondo effect. 
The difference between the current case and other reported Kondo systems is that the hybridization occurs between surface states, suggesting two-dimensional (2D) $c$-$f$ hybridization. 
The nature of this $c$-$f$ hybridization as well as the origin of the $F$ band splitting dependence on $k$ is examined in more detail in the following section.


\begin{figure*}[p]
\includegraphics[width=150mm]{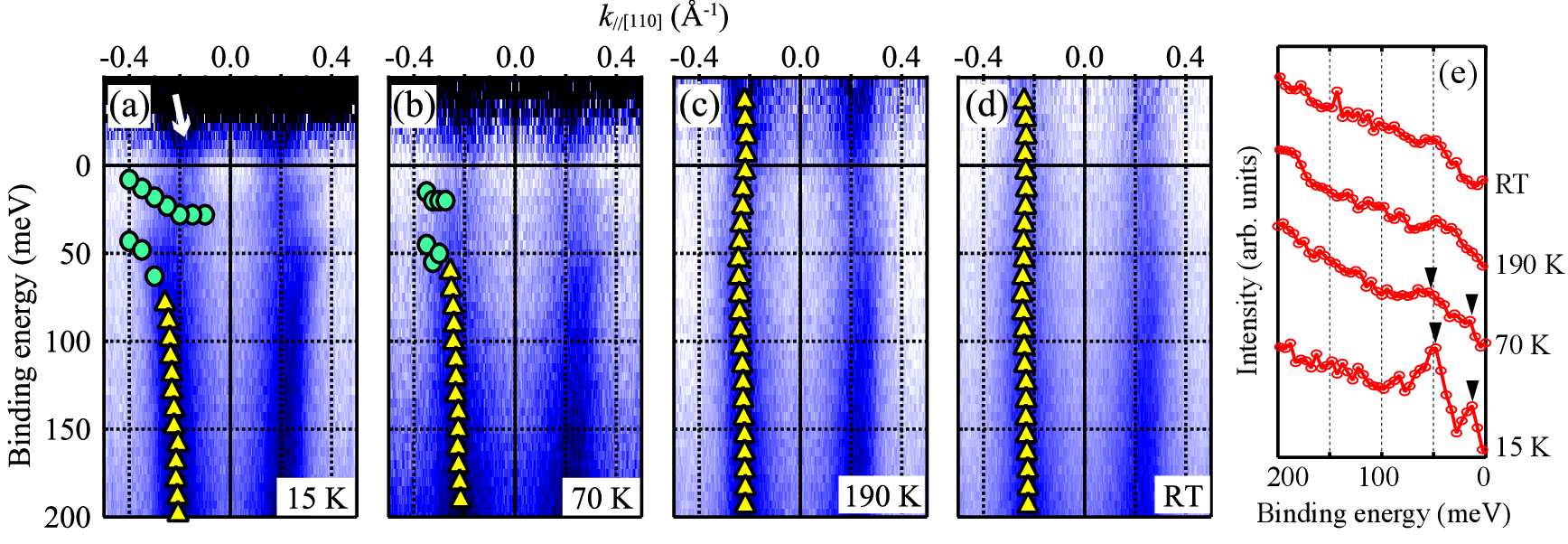}
\caption{\label{fig3}
(a-d) ARPES intensity plots near $E_{\rm F}$ at different temperatures divided by the Fermi distribution function at each sample temperature convolved with the instrumental resolution. All data were taken along [110] with $h\nu$ = 16.5 eV.
Circle (triangle) markers are the peak positions from ARPES momentum (energy) distribution curves.
The white arrow in (a) indicates a state lying at $E_{\rm F}$.
(e) ARPES energy distribution curves at $k_{//[110]}$ = 0.35 \AA$^{-1}$.
}
\end{figure*}

\subsection*{Temperature-dependent reconstruction of SS}

To examine the origin of the hybridization between the surface branches, we measured the temperature-dependent ARPES images at $h\nu$ = 16.5 eV as shown in Fig. 3.
The ARPES spectra were divided by the Fermi distribution function convolved with the instrumental resolution in order to observe the unoccupied states.
At 15 K, two separate bands are observed.
Compared with the ARPES image at Fig. 2 (c), the lower branch is composed of a continuous connection between $S$ and $F_{dn}$, namely the {\it c-f} hybridization band.
The other branch lies at a position between $F$ at $\bar{\Gamma}$ and $F_{up}$. However, these 4$f$ states at $\bar{\Gamma}$ and $k_{//[110]} >$ 0.4 \AA$^{-1}$ are not visible because the photoemission cross-section of 4$f$ electrons is negligibly small at this photon energy.
The 4$f$ states are observed as clear peaks at $k_{//[110]}$ = 0.2 to 0.3 \AA$^{-1}$ probably due to hybridization with the $S$ band, which should have Yb 5$d$ and/or B 2$sp$ character.
As a result of this hybridization, the $F$ branches at 0.2--0.3 \AA$^{-1}$, which are near the crossing point between $S$ and $F$, gain the orbital character of $S$ and become visible by ARPES at this photon energy.

At 70 K, the dispersion is almost the same as that at 15 K, but the separation between the 4$f$ branches is more diffuse than that at 15 K.
At 190 K, $S$ becomes a continuous metallic band across $E_{\rm F}$, and the 4$f$ states almost disappear.
On $S$, there is an undulation of the SS dispersion, which possibly a precursor of the hybridization with the 4$f$ states.
At room temperature, the undulation of $S$ is smaller but still visible.
The overall evolution of surface bands indicates that hybridization between the highly dispersive $S$ band and the almost-localized 4$f$ states depends on the temperature, which is driven by the Kondo effect.
This temperature-dependent behaviour, namely the disappearance of {\it c-f} hybridization at 190 K, agrees well with the previous reports about bulk states; the Kondo temperature of bulk YbB$_{12}$ is $\sim$220 K  \cite{Susaki96}, where the {\it c-f} gap size converges to zero.
In addition, the remaining undulation of $S$ is also consistent with the mid-infrared peak that still survives at room temperature \cite{Okamura05}.
These results suggest a close relationship between the ARPES results and the bulk electronic states.
However, Fig. 2 (b) indicates the 2D nature of the $S$ band.
Thus, the $F$ and $F_{up/dn}$ states intermixing with $S$ should also be localized in the surface/subsurface region.
One possible interpretation of this case is that the nature of the {\it c-f} hybridization on the YbB$_{12}$(001) surface is similar to that of the bulk bands.


\begin{figure*}[p]
\includegraphics[width=150mm]{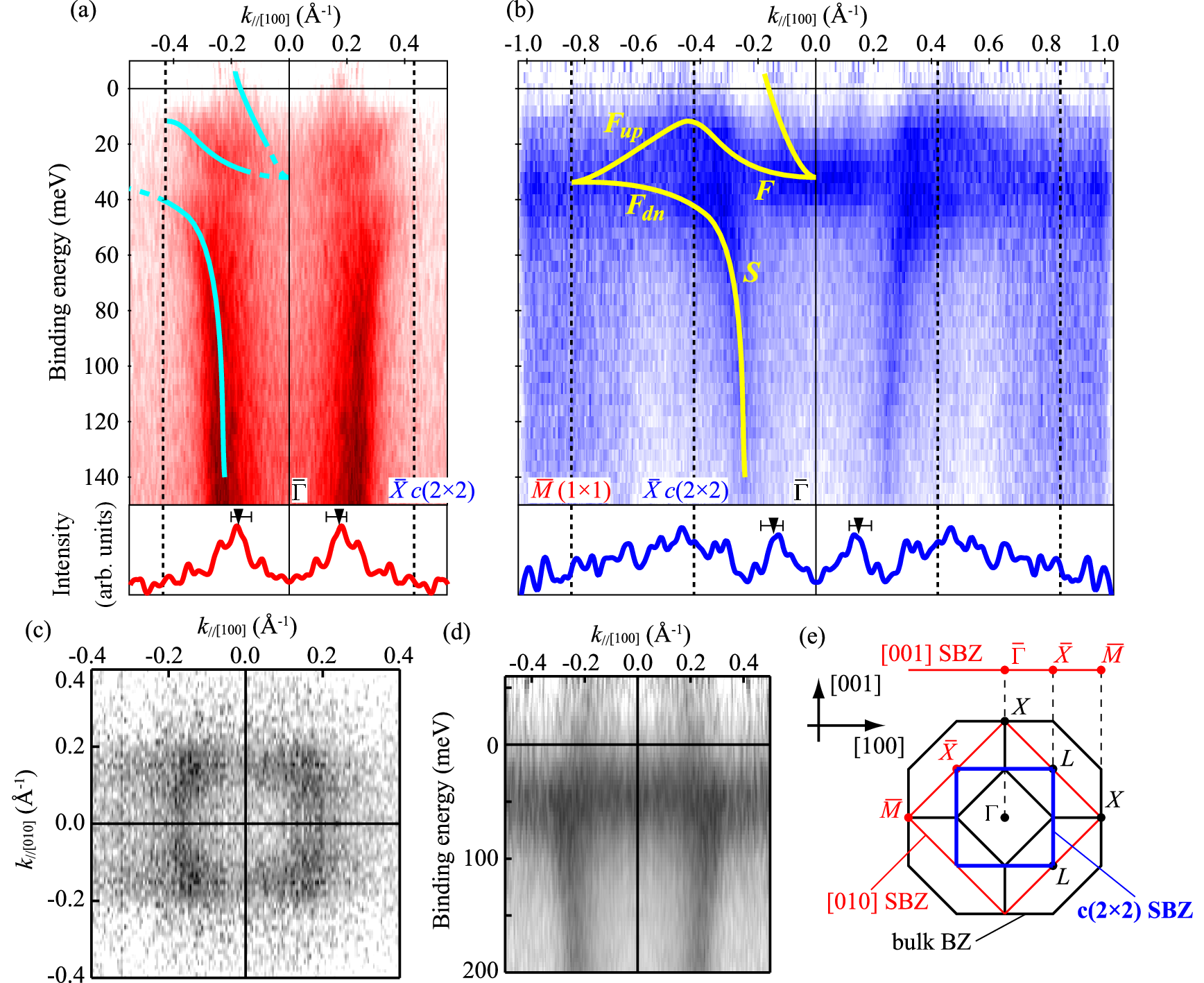}
\caption{\label{fig4}
(a, b) ARPES intensity plots along [100] near $E_{\rm F}$ taken with (a) 16.5 and (b) 53.5 eV photons and momentum distribution curves at $E_{\rm F}$ ($\pm$10 meV). The data is divided by the Fermi distribution function at the sample temperature (20 K for (a) and 14 K for (b)) convolved with the instrumental resolution. Lines are drawn as guides to the eye.
(c) Constant energy contour around $E_{\rm F}$ ($\pm$10 meV) taken with 16.5 eV photons at 14 K.
(d) ARPES image with 53.5 eV photons measured at room temperature.
(e) Schematic drawing of the 3D Brillouin zone of the YbB$_{12}$ single crystal and its projection onto the (001) surface Brillouin zone.
}
\end{figure*}

\subsection*{Discussion: the topological origin of SS}

In addition to the states discussed above, there is another state lying at $E_{\rm F}$ as indicated by the white arrow in Fig. 3 (a).
We focus on this state at low temperature in this section.
Figures 4 (a) and 4 (b) show the band dispersion along [100] taken at 16.5 and 53.5 eV, respectively.
At both photon energies, a new state crossing $E_{\rm F}$ at $k_{//[100]}\sim $ 0.18 \AA$^{-1}$ is clearly observed.
As shown in Fig. 4 (d), this state is absent at room temperature and is replaced with another SS with dispersion that is almost parallel to the bulk $c$ band; this SS is the same as that observed in Fig. 3 (d).
This state is degenerate with the 4$f$ ($F$) band at $\bar{\Gamma}$.
Moreover, this 4$f$ band disperses between $E_{\rm F}$ and 40 meV and becomes degenerate with the lower 4$f$ branch ($F_{dn}$) at $\sim$0.85 \AA$^{-1}$ ($\bar{M}$ of the (1$\times$1) SBZ, see Fig. 4(e)).
The complete SS dispersion exhibits a continuous connection across the Fermi level as well as the bulk $c$-$f$ hybridization gap, whose size is at most 100 meV \cite{Okamura05}, with twofold degeneracy at high symmetry points of the (1$\times$1) SBZ.
This metallic surface state can explain the remnant conduction path of YbB$_{12}$ observed at low temperatures \cite{Iga_Priv}.
Note that no states corresponding with the metallic state observed by ARPES were observed in the bulk bands \cite{Iga98, Okamura05}.

The dispersion of the surface state observed here agrees with the expected behaviour for topological surface states \cite{Kane10}: continuous dispersion across the bulk bandgap with Kramers degeneracy at the surface time-reversal invariant momentum (TRIM).
Indeed, the energy contour at $E_{\rm F}$ taken with 16.5 eV photons indicates a closed Fermi contour surrounding $\bar{\Gamma}$, one of the surface TRIM on (001), as shown in Fig. 4 (c).
This closed EC surrounding $\bar{\Gamma}$ agrees with the expected behaviour for topological surface states.

Note that this surface has $c$(2$\times$2) periodicity which might affect the surface band dispersion and make it difficult to determine the topological order of the material from the surface band dispersion.
However, the surface-state band observed here does not show any folding with respect to the $c$(2$\times$2) SBZ boundary as shown in Fig. 4 (b).
This behaviour suggests that the metallic SS is not derived from the $c$(2$\times$2) structure at the topmost surface but from the subsurface bulk-like region.
Such subsurface origin agrees with that of most of the previously discussed subsurface states that disperse parallel to the bulk bands as well as the topological surface states \cite{Eremeev11}.

\subsection*{Summary}

In summary, we discovered the new surface state on a clean surface of the YbB$_{12}$(001) Kondo insulator and surveyed its temperature-dependent reconstruction using ARPES.
The state was metallic and showed no dispersion along $k_z$, indicating its surface localization.
While the SS does not hybridize with the Yb $4f$ state lying immediately below the Fermi level ($E_{\rm F}$) at room temperature, strong hybridization occurred at low temperatures.
The reconstructed SS due to the $c$-$f$ hybridization at 20 K was metallic and dispersed continuously across the bulk bandgap of the Kondo insulator between $E_{\rm F}$ and the binding energy of $\sim$50 meV.
This SS behaviour at low temperatures agrees with the expected behaviour for SS on topological Kondo insulators.
Moreover, these results demonstrate the temperature-dependent holistic reconstruction of two-dimensional states localized on the surface of the Kondo insulator driven by Kondo effect.
Further study to identify the spin and orbital angular momentum polarization of the SS on YbB$_{12}$(001) is desirable to determine the topological order of YbB$_{12}$.

\subsection*{Methods}
The ARPES measurements were performed with synchrotron radiation at the CASSIOP\'EE beamline of the SOLEIL synchrotron, the BL7U beamline of UVSOR-III, and the BL-2A MUSASHI beamline of the Photon Factory.
The photon energies used in these measurements ranged from 15 to 1600 eV.
The incident photon is linearly polarized and the electric field of the photons lies in the incident plane (so-called $p$ polarization).
The photoelectron kinetic energy at $E_{\rm F}$ and the overall energy resolution of each ARPES setup ($\sim$20 meV for ARPES and $\sim$80 meV for wide-valence spectra shown in Figure 1 (d) and (e)) were calibrated using the Fermi edge of the photoelectron spectra from Ta foils attached to the sample.

Single crystalline YbB$_{12}$ was grown via the floating-zone method using an image furnace with four xenon lamps \cite{Iga98}.
The crystal was cut with a diamond saw along the (001) plane based on an $in-situ$ Laue pattern and was then polished in air until a mirror-like surface plane was obtained.
The polished YbB$_{12}$(001) crystal was heated to 1650 K in ultra-high vacuum chambers for $\sim$10 s.
For heating, a SiC wafer was underlaid below YbB$_{12}$(001); heating the SiC wafer with direct current simultaneously heats the attached YbB$_{12}$.

\subsection*{Acknowledgements}
We thank J. Kishi, Y. Takeno, and Y. Negoro for their support during general experiments.
For preliminary experiments to obtain a clean sample surface, we thank K. Imura, T. Hajiri, and T. Ito for their support.
We also acknowledge D. Ragonnet and F. Deschamps for their support during the experiments on the CASSIOP\'EE beamline at the SOLEIL synchrotron.
Part of the ARPES experiments were performed under UVSOR proposal Nos. 26-540 and 27-542 and Photon Factory proposal No. 2015G540.
This work was also supported by the JSPS Grant-in-Aid for Scientific Research Activity Start-up (Grant No. 26887024) and (B) (Grant No. 15H03676).

\subsection*{Author contributions}
K.H and Y.O. conducted the ARPES experiments with assistance from M.M., S.-I.I, K.T., J.R., P.R.F., F.B., A.T.-I., R.Y., M.K., K.H., and H.K.. F.I. grew the single-crystal samples. H.M. conducted the preliminary experiments to obtain a well-ordered clean sample surface. Y.O. and S.-I.K. wrote the text and were responsible for the overall direction of the research project.
All authors contributed to the scientific planning and discussions.

\end{document}